\RequirePackage{lineno}
\documentclass[floatfix,twocolumn,showpacs,preprintnumbers,amsmath,amssymb,prl,superscriptaddress]{revtex4-1} 
\usepackage{graphicx}
\usepackage{dcolumn}
\usepackage{bm}
\usepackage{soul}

\usepackage{color}
\begin{document}

\title{
\begin{flushright}
{\small  Version A4~~(\today)
 }
\end{flushright} 
Beam-Energy Dependence of Directed Flow of $\Lambda$, $\overline{\Lambda}$, $K^\pm$, $K^0_s$ and $\phi$ in Au+Au Collisions
}

\affiliation{AGH University of Science and Technology, FPACS, Cracow 30-059, Poland}
\affiliation{Argonne National Laboratory, Argonne, Illinois 60439}
\affiliation{Brookhaven National Laboratory, Upton, New York 11973}
\affiliation{University of California, Berkeley, California 94720}
\affiliation{University of California, Davis, California 95616}
\affiliation{University of California, Los Angeles, California 90095}
\affiliation{Central China Normal University, Wuhan, Hubei 430079}
\affiliation{University of Illinois at Chicago, Chicago, Illinois 60607}
\affiliation{Creighton University, Omaha, Nebraska 68178}
\affiliation{Czech Technical University in Prague, FNSPE, Prague, 115 19, Czech Republic}
\affiliation{Nuclear Physics Institute AS CR, 250 68 Prague, Czech Republic}
\affiliation{Frankfurt Institute for Advanced Studies FIAS, Frankfurt 60438, Germany}
\affiliation{Institute of Physics, Bhubaneswar 751005, India}
\affiliation{Indiana University, Bloomington, Indiana 47408}
\affiliation{Alikhanov Institute for Theoretical and Experimental Physics, Moscow 117218, Russia}
\affiliation{University of Jammu, Jammu 180001, India}
\affiliation{Joint Institute for Nuclear Research, Dubna, 141 980, Russia}
\affiliation{Kent State University, Kent, Ohio 44242}
\affiliation{University of Kentucky, Lexington, Kentucky 40506-0055}
\affiliation{Lamar University, Physics Department, Beaumont, Texas 77710}
\affiliation{Institute of Modern Physics, Chinese Academy of Sciences, Lanzhou, Gansu 730000}
\affiliation{Lawrence Berkeley National Laboratory, Berkeley, California 94720}
\affiliation{Lehigh University, Bethlehem, Pennsylvania 18015}
\affiliation{Max-Planck-Institut fur Physik, Munich 80805, Germany}
\affiliation{Michigan State University, East Lansing, Michigan 48824}
\affiliation{National Research Nuclear University MEPhI, Moscow 115409, Russia}
\affiliation{National Institute of Science Education and Research, HBNI, Jatni 752050, India}
\affiliation{National Cheng Kung University, Tainan 70101 }
\affiliation{Ohio State University, Columbus, Ohio 43210}
\affiliation{Institute of Nuclear Physics PAN, Cracow 31-342, Poland}
\affiliation{Panjab University, Chandigarh 160014, India}
\affiliation{Pennsylvania State University, University Park, Pennsylvania 16802}
\affiliation{Institute of High Energy Physics, Protvino 142281, Russia}
\affiliation{Purdue University, West Lafayette, Indiana 47907}
\affiliation{Pusan National University, Pusan 46241, Korea}
\affiliation{Rice University, Houston, Texas 77251}
\affiliation{Rutgers University, Piscataway, New Jersey 08854}
\affiliation{Universidade de Sao Paulo, Sao Paulo, Brazil, 05314-970}
\affiliation{University of Science and Technology of China, Hefei, Anhui 230026}
\affiliation{Shandong University, Jinan, Shandong 250100}
\affiliation{Shanghai Institute of Applied Physics, Chinese Academy of Sciences, Shanghai 201800}
\affiliation{State University of New York, Stony Brook, New York 11794}
\affiliation{Temple University, Philadelphia, Pennsylvania 19122}
\affiliation{Texas A\&M University, College Station, Texas 77843}
\affiliation{University of Texas, Austin, Texas 78712}
\affiliation{University of Houston, Houston, Texas 77204}
\affiliation{Tsinghua University, Beijing 100084}
\affiliation{University of Tsukuba, Tsukuba, Ibaraki, Japan,305-8571}
\affiliation{Southern Connecticut State University, New Haven, Connecticut 06515}
\affiliation{University of California, Riverside, California 92521}
\affiliation{United States Naval Academy, Annapolis, Maryland 21402}
\affiliation{Valparaiso University, Valparaiso, Indiana 46383}
\affiliation{Variable Energy Cyclotron Centre, Kolkata 700064, India}
\affiliation{Warsaw University of Technology, Warsaw 00-661, Poland}
\affiliation{Wayne State University, Detroit, Michigan 48201}
\affiliation{World Laboratory for Cosmology and Particle Physics (WLCAPP), Cairo 11571, Egypt}
\affiliation{Yale University, New Haven, Connecticut 06520}

\author{L.~Adamczyk}\affiliation{AGH University of Science and Technology, FPACS, Cracow 30-059, Poland}
\author{J.~R.~Adams}\affiliation{Ohio State University, Columbus, Ohio 43210}
\author{J.~K.~Adkins}\affiliation{University of Kentucky, Lexington, Kentucky 40506-0055}
\author{G.~Agakishiev}\affiliation{Joint Institute for Nuclear Research, Dubna, 141 980, Russia}
\author{M.~M.~Aggarwal}\affiliation{Panjab University, Chandigarh 160014, India}
\author{Z.~Ahammed}\affiliation{Variable Energy Cyclotron Centre, Kolkata 700064, India}
\author{N.~N.~Ajitanand}\affiliation{State University of New York, Stony Brook, New York 11794}
\author{I.~Alekseev}\affiliation{Alikhanov Institute for Theoretical and Experimental Physics, Moscow 117218, Russia}\affiliation{National Research Nuclear University MEPhI, Moscow 115409, Russia}
\author{D.~M.~Anderson}\affiliation{Texas A\&M University, College Station, Texas 77843}
\author{R.~Aoyama}\affiliation{University of Tsukuba, Tsukuba, Ibaraki, Japan,305-8571}
% ...

\collaboration{STAR Collaboration}\noaffiliation

\begin{abstract}
Rapidity-odd directed flow measurements at midrapidity are presented for $\Lambda$, $\overline{\Lambda}$, $K^\pm$, $K^0_s$ and $\phi$ at $\sqrt{s_{NN}} =$ 7.7, 11.5, 14.5, 19.6, 27, 39, 62.4 and 200 GeV in Au+Au collisions recorded by the Solenoidal Tracker detector at the Relativistic Heavy Ion Collider. These measurements greatly expand the scope of data available to constrain models with differing prescriptions for the equation of state of quantum chromodynamics. Results show good sensitivity for testing a picture where flow is assumed to be imposed before hadron formation and the observed particles are assumed to form via coalescence of constituent quarks. The pattern of departure from a coalescence-inspired sum-rule can be a valuable new tool for probing the collision dynamics. 
\end{abstract}

\maketitle
\realpagewiselinenumbers
\setlength\linenumbersep{0.10cm}

Rapidity-odd directed flow, $v_1^{\rm odd}(y)$, is the first harmonic coefficient in the Fourier expansion of the final-state azimuthal distribution relative to the collision reaction plane \cite{methods}, and describes a collective sideward motion of emitted particles. The rapidity-even component $v_1^{\rm even}(y)$ \cite{Luzum} is unrelated to the reaction plane in mass-symmetric collisions, and arises from event-by-event fluctuations in the initial nuclei. Hereafter, $v_1(y)$ implicitly refers to the odd component. Both hydrodynamic \cite{hydro} and nuclear transport \cite{urqmd} models indicate that $v_1(y)$ is sensitive to details of the expansion during the early stages of the collision fireball \cite{prequilibrium,Heinz}. To integrate over the rapidity dependence, it is common practice to present $dv_1/dy$ near midrapidity, as in the Solenoidal Tracker at RHIC (STAR) measurements for protons, antiprotons and pions in Au+Au collisions at $\sqrt{s_{NN}} =$ 7.7 to 200 GeV.  Both protons and net protons show a minimum in $dv_1/dy$ near $\sqrt{s_{NN}}$ of 10 to 20 GeV \cite{STAR-BESv1}.  Based on hydrodynamic calculations \cite{Rischke,Stoecker}, a minimum in directed flow has been proposed as a signature of a first-order phase transition between hadronic matter and quark-gluon plasma (QGP). 

There have been several recent $v_1(y)$ model calculations with various assumed quantum chromodynamics (QCD) equations of state \cite{hybrid,phsd,phsd2,3FD,JAM,JAM2}. The assumption of purely hadronic physics is disfavored, but there is no consensus on whether STAR measurements \cite{STAR-BESv1} favor a crossover or first-order phase transition. Models do not produce any $dv_1/dy$ minimum over the observed energy range \cite{hybrid,phsd,phsd2,3FD,JAM,LaszloHorst}, with the exception of one case where a minimum was calculated near one-third the energy of the measured minimum \cite{JAM2}. Moreover, predicted $v_1$ is strongly sensitive to model details unrelated to the assumed equation of state \cite{v1review}. Thus, further progress in models is needed for a definitive interpretation. 

Number-of-constituent-quark (NCQ) scaling \cite{STARwhitepaper} (whereby elliptic flow ($v_2$) behaves as if imposed at the level of deconfined constituent quarks) is an example of coalescence behavior among quarks.  There is a history of coalescence observations in heavy-ion collisions, in the formation of nuclei \cite{EarlyTheory,BevalacComposites,EOSflow,STARanti-alpha} as well as in the hadronization of quarks.  The interplay between NCQ scaling and the transport of initial-state $u$ and $d$ quarks towards midrapidity during the collision offers possibilities for new insights \cite{DunlopLisaSorensen}.  However, this physics remains poorly understood \cite{BES-I,BES-II}, and these considerations motivate $v_n$ versus $\sqrt{s_{NN}}$ measurements encompassing as many particle species as possible. 

\begin{figure*}[!ht]
\centering
\centerline{\includegraphics[scale=0.85]{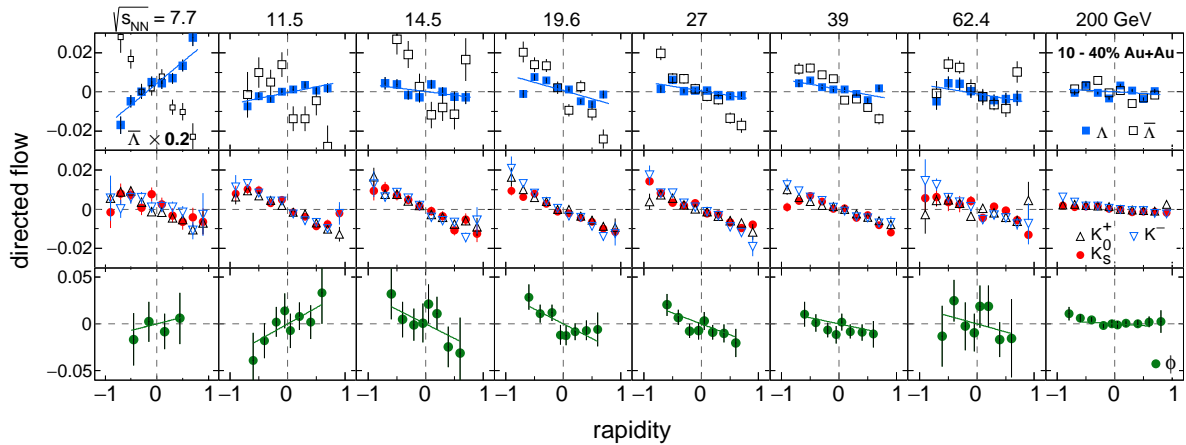}}
\caption{(Color online) Directed flow as a function of rapidity for the six indicated particle species in 10-40\% central Au+Au collisions at $\sqrt{s_{NN}} = 7.7$ to 200 GeV. The error bars include statistical uncertainties only; systematic errors are presented in Fig.~\ref{fig2}.  The two upper panel rows use the same $v_1$ scale with the exception of $\overline{\Lambda}$ at $\sqrt{s_{NN}} = 7.7$ GeV, where $v_1$ magnitudes are exceptionally large and require the measurements to be divided by five. Examples of linear fits to $v_1(y)$ are shown in the case of $\Lambda$ and $\phi$. }
\label{fig1}
\end{figure*}

We report the first measurements of directed flow versus rapidity for $\Lambda$, $\overline{\Lambda}$, $\phi$, $K^\pm$ and $K^0_s$ in Au+Au collisions at eight beam energies $\sqrt{s_{NN}}= 7.7$, 11.5, 14.5, 19.6, 27, 39, 62.4 and 200 GeV, where the analyzed samples contain 4, 12, 20, 36, 70, 130, 50 and 250 million minimum-bias-trigger events, respectively.  These data from the STAR~\cite{starnim} located at Brookhaven National Laboratory were recorded in 2010, 2011 and 2014.  The STAR Time Projection Chamber (TPC) \cite{startpc} was used for charged-particle tracking within pseudorapidity $|\eta| < 1$.  The centrality was determined from the number of charged particles within $|\eta| < 0.5$.  For determination of the event plane \cite{methods}, two Beam-Beam Counters (BBC, pseudorapidity coverage $3.3 < |\eta| < 5.0$ for inner tiles) were used at $\sqrt{s_{NN}} \leq 39$ GeV \cite{BBC, STAR-BESv1}, while the STAR Zero-Degree Calorimeter Shower-Maximum Detectors (ZDC-SMD, $|\eta| > 6.3$) were used at $\sqrt{s_{NN}} = 62.4$ and 200 GeV\cite{GangThesis, v1-62, v1-4systems, pidv1_200, STAR-BESv1}. 

We require the primary vertex position of each event along the beam direction to lie within 70 cm of the center of the detector at $\sqrt{s_{NN}} = 7.7$ GeV, within 50 cm at 11.5 to 27 GeV, and within 40 cm at 39 to 200 GeV.  Tracks are required to have transverse momenta $p_T > 0.2$ GeV$/c$, have a distance of closest approach to the primary vertex of less than 3 cm, have at least 15 space points in the TPC acceptance $(|\eta| < 1.0)$, and have a ratio of the number of measured space points to the maximum possible number of space points greater than 0.52. This last requirement prevents double-counting of a particle due to track splitting. Charged kaons with $p_T > 0.2$ GeV$/c$ and momentum $< 1.6$ GeV$/c$ are identified based on energy loss in the TPC and time-of-flight information from the TOF detector \cite{TOF}. $\Lambda$, $\overline{\Lambda}$ and $K^0_s$ within $0.2 < p_T < 5.0$ GeV$/c$ and $\phi$ within $0.15 < p_T < 10.0$ GeV$/c$ are selected by standard V0 topology cuts using the invariant mass technique with mixed-event background subtraction \cite{STAR-BESv2}.  

\begin{figure}[!htb]
\begin{center}
\includegraphics[scale=0.42]{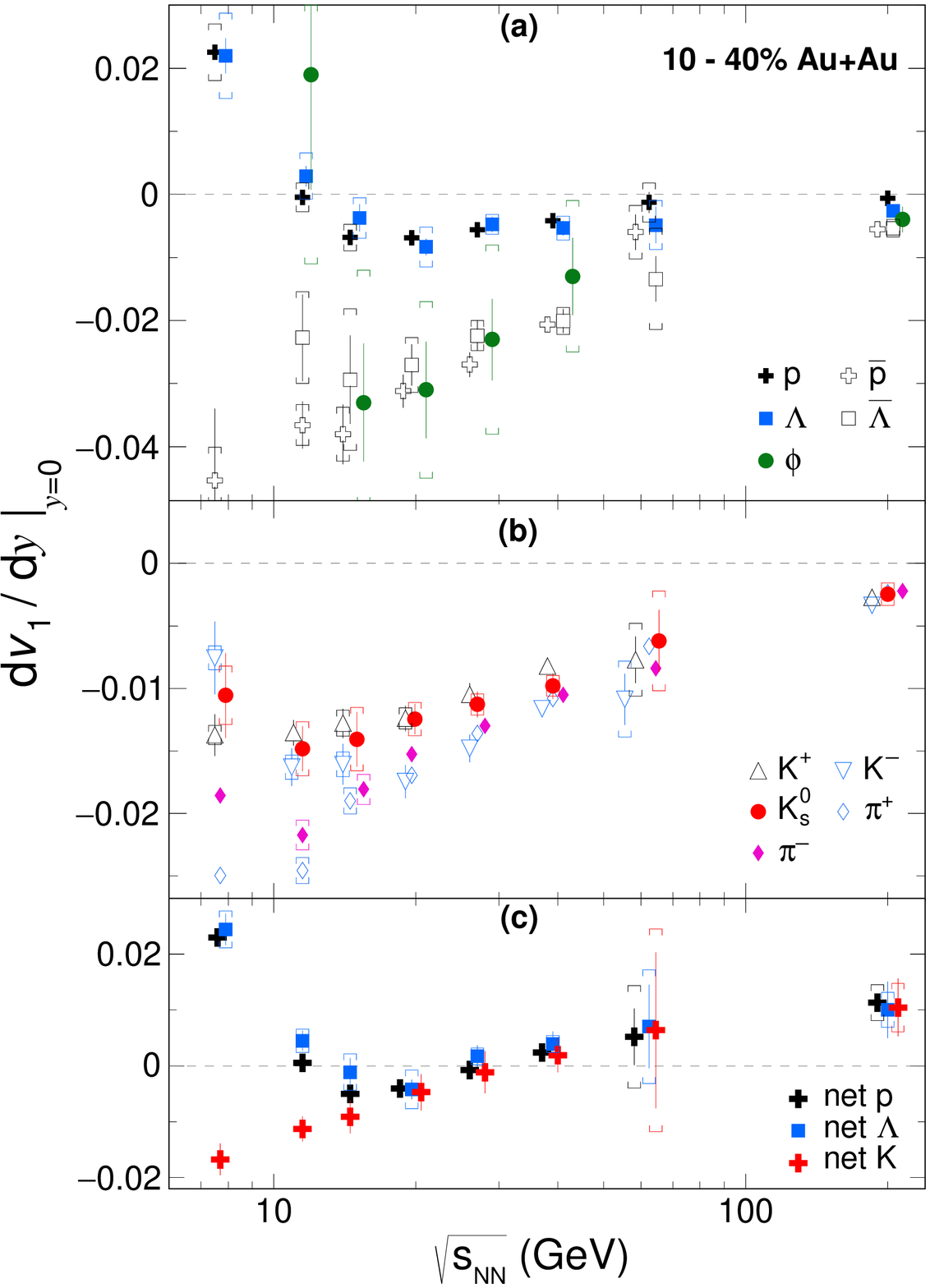}
\caption{(Color online) 
Directed flow slope ($dv_1/dy$) versus beam energy for intermediate-centrality (10-40\%) Au+Au collisions.  Panel (a) presents heavy species: $\Lambda$, $\overline{\Lambda}$, protons, antiprotons and $\phi$, while panel (b) presents  $K^\pm$, $K^0_s$ and $\pi^\pm$.  Note that $dv_1/dy$ for $\overline{\Lambda}$ at $\sqrt{s_{NN}} = 7.7$ GeV is $-0.128 \pm 0.022$ (stat) $\pm 0.026$ (sys), which is far below the bottom of the plotted scale. The $\phi$-meson result at $\sqrt{s_{NN}} = 62.4$ GeV has a large uncertainty and is not plotted. Panel (c) presents net protons, net $\Lambda$s, and net kaons. The bars are statistical errors, while the caps are systematic uncertainties. Data points are staggered horizontally to improve visibility.
}
\label{fig2}
\end{center}
\end{figure}

\begin{figure}[!htb]
\begin{center}
\includegraphics[scale=0.42]{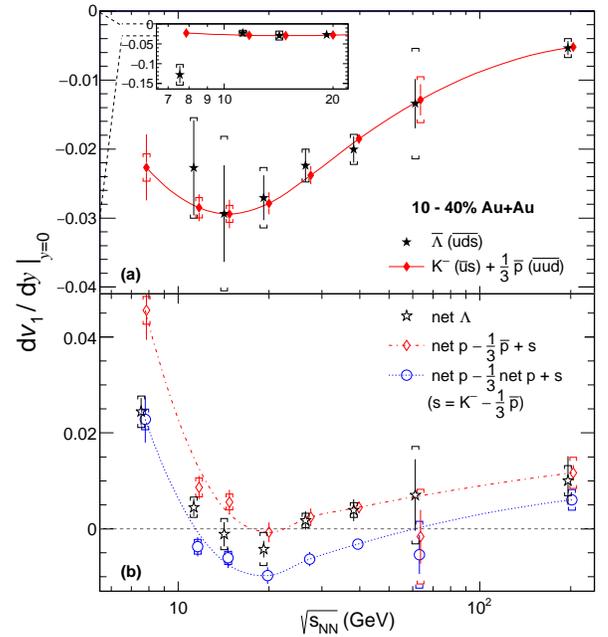}
\caption{(Color online) Directed flow slope ($dv_1/dy$) versus $\sqrt{s_{NN}}$ for intermediate centralities (10-40\%). Panel (a) compares the observed $\overline{\Lambda}$ slope with the prediction of the coalescence sum rule for produced quarks. The inset shows the same comparison where the vertical scale is zoomed-out; this allows the observed flow for the lowest energy ($\sqrt{s_{NN}} = 7.7$ GeV) to be seen.  Panel (b) presents two further sum-rule tests, based on comparisons with net-$\Lambda$ measurements. The solid and dotted lines are smooth curves to guide the eye.} 
\label{fig3}
\end{center}
\end{figure}

Systematic uncertainties arising from event-plane estimation in essentially the same $v_1$ analysis for different species are discussed elsewhere~\cite{STAR-BESv1}. Non-flow is a source of possible systematic error that refers to azimuthal correlations unrelated to the reaction plane orientation, arising from resonances, jets, strings, quantum statistics and final-state interactions like Coulomb effects.  Possible non-flow effects are reduced due to the sizable pseudorapidity gap between the TPC and the BBC or ZDC-SMD detectors~\cite{methods}. We have studied the sensitivity of $dv_1/dy$ to all experimental cuts and selections, for both events and tracks, and inferred systematic errors are plotted in Figs. \ref{fig2} and \ref{fig3}. 

Figure \ref{fig1} presents $v_1(y)$ at 10-40\% centrality for $K^\pm$, $K^0_s$, $\phi$, $\Lambda$ and $\overline{\Lambda}$.  These measurements complement the corresponding published information for protons, antiprotons and charged pions \cite{STAR-BESv1}. In the referenced $v_1$ study, the overall strength of the directed flow signal near midrapidity was characterized by the linear term $F$ in a fit of the form $v_1(y) = Fy + F_3 y^3$ \cite{STAR-BESv1}. This cubic fit reduces sensitivity to the rapidity range over which the fit is performed, but becomes unstable for low statistics, as is now the case for $\phi$ and $\overline{\Lambda}$, and to a lesser extent for $\Lambda$. Accordingly, the present analysis uses a linear fit for all particle species at all beam energies. The fit is over $|y| < 0.6$ for $\phi$ and over $|y| < 0.8$ for all other species. It is evident from Fig.~\ref{fig1} that within errors, the plotted species have a near-linear $v_1(y)$ over the acceptance of the STAR detector. However, protons \cite{STAR-BESv1} show systematic deviations from linearity and hence the proton $dv_1/dy |_{y=0}$ is marginally affected by changing the fit method. Hereafter, $dv_1/dy$ refers to the slope obtained from the above linear fits.

Directed flow slope $dv_1/dy$ versus beam energy for $p$, $\overline{p}$, $\Lambda$, $\overline{\Lambda}$, $\phi$, $K^\pm$, $K^0_s$, and $\pi^\pm$ is presented in Figs.~\ref{fig2}(a) and \ref{fig2}(b). The proton and pion points in Fig. \ref{fig2} differ slightly from those in Ref.~\cite{STAR-BESv1} in that a new measurement at $\sqrt{s_{NN}}= 14.5$ GeV has been added, and the slope is now based on a linear fit. We note four empirical patterns based on Figs. \ref{fig2}(a) and (b). First, $dv_1/dy$ for $\Lambda$ and $p$ agree within errors, and the $\Lambda$ slope changes sign in the same region as protons (near $\sqrt{s_{NN}} = 11.5$ GeV).  However, the $\Lambda$ errors are not small enough to determine whether the minimum observed in the proton slope near $\sqrt{s_{NN}} = 15$ to 20 GeV also occurs for $\Lambda$. Second, $dv_1/dy$ for $K^+$ and $K^-$ are both negative at all energies and are close to each other except at the lowest energy, while $dv_1/dy$ for $K^0_s$ is everywhere consistent within errors with the average of $K^+$ and $K^-$.  It was found previously that $dv_1/dy$ for $\pi^+$ and $\pi^-$ is likewise close over these energies and is always negative.  Third, the slope for $\overline{\Lambda}$ is negative throughout and is consistent within errors with $\overline{p}$ \cite{STAR-BESv1}.  Fourth, at $\sqrt{s_{NN}} = 14.5$ GeV and above, the $\phi$ slope has much larger magnitude than other mesons (pions and kaons) and is close to $\overline{p}$ and $\overline{\Lambda}$. At $\sqrt{s_{NN}} = 11.5$ GeV, the $dv_1/dy$ for $\phi$ increases steeply, although the statistical significance of the increase is poor. The $\phi$-meson $v_1$ statistics are too marginal to permit a reliable determination of $dv_1/dy$ at $\sqrt{s_{NN}} = 7.7$ GeV. 

Particles like $p$, $\Lambda$ and $K^+$ receive more contributions from transported quarks ($u$ and $d$ from the initial-state nuclei) than their antiparticles \cite{DunlopLisaSorensen}. ``Net particle" represents the excess yield of a particle species over its antiparticle.  In order to enhance the contribution of transported quarks relative to those produced in the collision, we define $v_{1\,{\rm net\,}p}$ based on expressing $v_1(y)$ for all protons as 
$$v_{1\,p} = r(y) \,v_{1\,\overline{p}} + [1 - r(y)]\, v_{1\,{\rm net\,}p},$$
where $r(y)$ is the ratio of observed $\overline{p}$ to $p$ yield at each beam energy.  Corrections of $r(y)$ for reconstruction inefficiency and backgrounds were found to have a negligible effect on the net-proton $dv_1/dy$ and have not been applied.  
Figure \ref{fig2}(c) presents net-proton $dv_1/dy$, and also includes net-$\Lambda$ and net-kaon $dv_1/dy$,  
defined similarly, except $\overline{p}\,(p)$ becomes $\overline{\Lambda}\,(\Lambda)$ and $K^-(K^+)$, respectively.

The ten particle species available in the present analysis allow a more detailed investigation of constituent-quark $v_1$ than was possible in Ref.~\cite{STAR-BESv1}.  We are now in a position to test a set of assumptions, namely that $v_1$ is imposed at the pre-hadronic stage, that specific types of quark have the same directed flow, and that the detected hadrons are formed via coalescence \cite{STARwhitepaper, DunlopLisaSorensen}.  In a scenario where deconfined quarks have already acquired azimuthal anisotropy, and in the limit of small azimuthal anisotropy coefficients $v_n$, coalescence leads to the $v_n$ of the resulting mesons or baryons being the summed $v_n$ of their constituent quarks \cite{FriesGrecoSorensen, DunlopLisaSorensen}.  We call this assumption the coalescence sum rule.  NCQ scaling in turn follows from the coalescence sum rule \cite{DunlopLisaSorensen}. Note that no weights are involved in coalescence sum rule $v_1$ calculations, unlike the case of $v_1$ for net particles.  

Antiprotons and $\overline{\Lambda}$s are seen to have similar $v_1(y)$, and it is noteworthy that these species are composed of three constituent quarks all produced in the collision, as opposed to being composed of $u$ or $d$ quarks which could be either transported from the initial nuclei or produced.  To test the coalescence sum rule in a straightforward case where all quarks are known to be produced, Fig.~\ref{fig3}(a) compares the observed $dv_1/dy$ for $\overline{\Lambda}$($\overline{uds}$) with the calculation for $K^-(\overline{u}s) + {1 \over 3}\overline{p}\,(\overline{uud})$. This calculation is based on the coalescence sum rule combined with the assumption that $s$ and $\overline{s}$ quarks have the same flow, and that $\overline{u}$ and $\overline{d}$ have the same flow.  The factor $1 \over 3$ arises from assuming that all $\overline{u}$ and $\overline{d}$ quarks contribute the same flow.  Close agreement is observed at $\sqrt{s_{NN}}= 11.5$ to 200 GeV.  The inset in Fig.~\ref{fig3}(a) presents the same comparison, but with a much coarser vertical scale. The observed sharp breakdown of agreement at $\sqrt{s_{NN}} = 7.7$ GeV implies that one or more of the above-mentioned assumptions no longer hold below 11.5 GeV.  A similar decrease in the produced-quark $v_2$ has been observed in the same energy region \cite{STAR-BESv2,STAR-BESv2-2016}.  

Next, we turn our attention to the less straightforward case of coalescence involving $u$ and $d$ quarks. We expect $v_1$ to be quite different for transported and produced quarks, which are difficult to distinguish in general.  However, in the limit of low $\sqrt{s_{NN}}$, most $u$ and $d$ quarks are presumably transported, while in the limit of high $\sqrt{s_{NN}}$, most $u$ and $d$ are produced. In Fig.~\ref{fig3}(b), we test two coalescence sum rule scenarios which are expected to bracket the observed $dv_1/dy$ for a baryon containing transported quarks.  The fraction of transported quarks among the constituent quarks of net particles is larger than in particles roughly in proportion to $N_{\rm particle}/N_{\rm net\, particle}$ \cite{NetRatio}, and therefore we employ net-$\Lambda$ and net-proton $v_1$ in these tests. 

Figure \ref{fig3}(b) presents the observed $dv_1/dy$ for net $\Lambda$($uds$). The first compared calculation (red diamond markers) consists of net protons ($uud$) minus $\overline{u}$ plus $s$, where $\overline{u}$ is estimated from ${1 \over 3} \overline{p}$, while the $s$ quark flow is obtained from $K^-(\overline{u}s) - {1 \over 3} \overline{p}\,(\overline{uud})$. There is no corresponding clear-cut expression for transported $u$ and $d$ quarks. Here, it is assumed that a produced $u$ quark in net $p$ is replaced with an $s$ quark.  This sum-rule calculation agrees closely with the net-$\Lambda$ measurement at $\sqrt{s_{NN}}= 19.6$ GeV and above, remains moderately close at 14.5 and 11.5 GeV, and deviates significantly only at 7.7 GeV.  The fraction of transported quarks among the constituent quarks of net protons increases with decreasing beam energy, and there is an increasing departure from the assumption that a produced $u$ quark is removed by keeping the term (net $p - {1 \over 3}\overline{p}$). 

The second coalescence calculation in Fig.~\ref{fig3}(b) corresponds to ${2 \over 3}$ net proton plus $s$ (blue circle markers). In this case, it is assumed that the constituent quarks of net protons are dominated by transported quarks in the limit of low beam energy, and that one of the transported quarks is replaced by $s$. This approximation breaks down as the beam energy increases, and there is disagreement between the black stars and blue circles above $\sqrt{s_{NN}} = 7.7$ GeV.  At $\sqrt{s_{NN}} = 62.4$ and 200 GeV, the size of errors and the closeness of the two sum rule calculations are such that no discrimination between the two scenarios is possible. 

In summary, we report $v_1(y)$ for $\Lambda$, $\overline{\Lambda}$, $\phi$, $K^\pm$ and $K^0_s$ at eight $\sqrt{s_{NN}}$ values spanning 7.7 to 200 GeV.  We focus on $dv_1/dy$ at midrapidity for 10-40\% centrality.  The directed flow slopes as a function of beam energy for protons and $\Lambda$s agree within errors, and change sign near 11.5 GeV.  Antiprotons, $\overline{\Lambda}$, kaons and pions have negative $dv_1/dy$ throughout the studied energy range. Net-particle $dv_1/dy$ for $p$, $\Lambda$ and $K$ agree at and above $\sqrt{s_{NN}}=14.5$ GeV, but net kaons increasingly diverge at 11.5 and 7.7 GeV. Overall, several features of the data undergo a prominent change near the lower beam energies. Some of the measurements are consistent with the observed particles having formed via coalescence of constituent quarks. The observed pattern of scaling behavior for produced quarks at and above 11.5 GeV, with a breakdown at 7.7 GeV, requires further study.  One hypothesis is that there is a turn-off below 11.5 GeV of the conditions for quark coalescence sum rule behavior, or a breakdown of the assumption that $s$ and $\overline{s}$ quarks have the same flow, or a breakdown of the assumption that $\overline{u}$ and $\overline{d}$ have the same flow. The energy-dependent measurements reported here will be enhanced after STAR acquires greatly increased statistics using upgraded detectors in Phase-II of the RHIC Beam Energy Scan \cite{BES-II}. 

\begin{acknowledgments}
We thank the RHIC Operations Group and RCF at BNL, the NERSC Center at LBNL, and the Open Science Grid consortium for providing resources and support. This work was supported in part by the Office of Nuclear Physics within the U.S. DOE Office of Science, the U.S. National Science Foundation, the Ministry of Education and Science of the Russian Federation, National Natural Science Foundation of China, Chinese Academy of Science, the Ministry of Science and Technology of China and the Chinese Ministry of Education, the National Research Foundation of Korea, GA and MSMT of the Czech Republic, Department of Atomic Energy and Department of Science and Technology of the Government of India; the National Science Centre of Poland, National Research Foundation, the Ministry of Science, Education and Sports of the Republic of Croatia, RosAtom of Russia and German Bundesministerium f\"ur Bildung, Wissenschaft, Forschung and Technologie (BMBF) and the Helmholtz Association.
\end{acknowledgments}


\begin{thebibliography}{00}
\bibitem{NetRatio}
This ratio is based on the assumption that all the transported quarks are contained in net particles.
\end{thebibliography}
\end{document}